\documentclass[aps,prd,twocolumn]{revtex4}
\usepackage{graphicx}
\usepackage{amssymb}
\usepackage{epsfig}
\usepackage{bm}
\usepackage{dcolumn}

\newcommand{\grado}{^{\circ}}
\def\beq{ \begin{equation}}
\def\eeq{\end{equation} }
\def\bea{\begin{eqnarray}}
\def\eea{\end{eqnarray}}

\newcommand{\etal}{et al.}

\begin{document}

\title{Dark Matter annihilation in Draco: new considerations of the expected gamma flux}

\author{M. A. S\'anchez-Conde$^1$, F. Prada$^1$, E. L. {\L}okas$^2$, M. E. G\'omez$^3$, R. Wojtak$^2$ and M. Moles$^1$} 
\affiliation{$^1$ Instituto de Astrof\'isica de Andaluc\'ia (CSIC), E-18008, Granada, Spain} 
\affiliation{$^2$ Nicolaus Copernicus Astronomical Centre, Bartycka 18, 00-716 Warsaw, Poland}
\affiliation{$^3$ Departamento de F\'isica Aplicada, Facultad de Ciencias Experimentales, Universidad de Huelva, 21071 Huelva, Spain}

\begin{abstract}
  A new revision of the gamma flux that we expect to detect in Imaging
  Atmospheric Cherenkov Telescopes (IACTs) from neutralino
  annihilation in the Draco dSph is presented in the context of the
  minimal supersymmetric standard models (MSSM) compatible with the
  present phenomenological and cosmological constraints, and using the
  dark matter (DM) density profiles compatible with the latest
  observations. This revision takes also into account the important
  effect of the Point Spread Function (PSF) of the telescope, and is
  valid not only for Draco but also for any other DM target. We show
  that this effect is crucial in the way we will observe and interpret
  a possible signal detection. Finally, we discuss the prospects to
  detect a possible gamma signal from Draco for current or planned
  $\gamma$-ray experiments, i.e. MAGIC, GLAST and GAW. Even with the
  large astrophysical and particle physics uncertainties we find that
  the chances to detect a neutralino annihilation signal in Draco seem
  to be very scarce for current experiments. However, the prospects
  for future IACTs with upgraded performances (especially lower
  threshold energies and higher sensitivities) such as those offered
  by the CTA project, might be substantially better.
\end{abstract}


\maketitle

\section{Introduction}

Nowadays, it is generally believed that only a small fraction of the
matter in the Universe is luminous. In the Cold Dark Matter (CDM)
cosmological scenario around one third of the
``dark'' side of the Universe is supposed to be composed of
weakly interacting massive particles (WIMPs). although other possible
candidates like axinos or gravitinos are not excluded (see
\cite{bertone} for a recent review). The Standard Model can not
provide a suitable explanation to the dark matter (DM)
problem. However, its supersymmetric extension (SUSY) provides a
natural candidate for DM in the form of a stable uncharged Majorana
fermion, called neutralino, which constitutes also one of the most
suitable candidates according to the current cosmological
constraints. At present, large effort is being carried out to detect
this SUSY DM by different methods \cite{Munoz:2003gx}. In the case of
the new Imaging Atmospheric Cherenkov Telescopes (IACTs), the searches
are based on the detectability of gamma rays coming from the
annihilation of the SUSY DM particles. IACTs in operation like MAGIC
\cite{magic} or HESS \cite{hess}, or satellites-based experiments like the upcoming GLAST
satellite \cite{glast}, will play a very important role in these DM
searches.

A relevant question concerning the search of SUSY DM is where to
search for the annihilation gamma ray signal. Due to the fact that the
gamma flux is proportional to the square of the DM density, we will
need to point the telescope to places where we expect to find a high
concentration of dark matter. In principle, the best option seems to
be the Galactic Centre (GC), since it satisfies this condition and it
is also very near compared to other potential targets. However, the GC
is a very crowded region, which makes it difficult to discriminate
between a possible $\gamma$-ray signal due to neutralino annihilation
and other astrophysical sources. Whipple \cite{whipple}, Cangaroo
\cite{cangaroo}, and specially HESS \cite{hess04} and MAGIC
\cite{magic06} have already carried out detailed observations of the
GC and all of them reported a gamma point-like source at the Sag A*
location. However, if this signal was interpreted as fully due to DM
annihilation, it would correspond to a very massive neutralino very
difficult to fit within the WMAP cosmology \cite{wmap} in the
preferred SUSY framework \cite{bergstrom05} (although an alternative
scenario with multi-TeV neutralinos compatible with WMAP is still
possible, see \cite{multiTeV}). Furthermore, an extended emission was
also discovered in the GC area, but it correlates very well with
already known dense molecular clouds \cite{hess06}. Recently, new HESS
data on the GC have been published and a reanalysis has been carried
out by the HESS collaboration \cite{hess06GC}. In this work, they
especially explore the possibility that some portion of the detected
signal is due to neutralino annihilation. According to their results,
at the moment it is not possible to exclude a DM component hidden
under a non-DM power-law spectrum due to an astrophysical source.

There are also other possible targets with high dark matter density in
relative proximity from us, e.g. the Andromeda galaxy, the dwarf spheroidal (dSph)
galaxies - most of them satellites of the Milky Way- or even massive
clusters of galaxies (e.g. Virgo). DSph galaxies represent a good
option, since they are not plagued by the problems of
the GC, they are dark matter dominated systems with very high
mass to light ratios, and at least six of them are nearer than 100 kpc
from the GC (Draco, LMC, SMC, CMa, UMi and Sagittarius).

Concerning DM detection, there are two unequivocal signatures that
make sure that the $\gamma$-ray signal is due to neutralino
annihilation: the spectrum of the source and its spatial extension. Certainly, the spectrum shape is crucial to state that the gamma source is due to DM annihilation, not only to discriminate it from the astrophysical backgrounds (e.g. hadronic, electronic and diffuse for IACTs), but also from any other astrophysical sources. The keypoint here is that the spectrum of a DM annihilation source is supposed to be in concordance with that expected from the models of particle physics. Indeed, the exact annihilation spectra of MSSM neutralinos depend on the gaugino/higgsino mixing, but all of them show a continuum curved spectrum up to the mass of the DM particle and possibly faint $\gamma$-lines superimposed \cite{bergstrom00}. This spectrum will be very different from those measured in the case of an astrophysical source \cite{hess06GC}. As for the spatial extension of the source, in this case the important feature is that it should be extended and diffuse, showing also a characteristic shape of the gamma flux profile. Nevertheless, we must note that if we use an instrument that does not have a spatial resolution good enough compared to the extension of the source, we might see only a point-like source instead of a diffuse or extended one. This means that, although we might reach high enough sensitivity for a successful detection, we would not be completely sure whether our signal is due to DM annihilation or not. Therefore, it is clear that it is really important to resolve the source so we can conclude that it can be
interpreted as neutralino annihilation.

In this work, we first calculate the gamma ray flux profiles expected
in a typical IACT due to neutralino annihilation in the Draco dSph,
which represents in principle a good candidate to search for
DM. Draco, located at 80 kpc, is one of the dwarfs with many
observational constraints, which has helped to determine better its DM
density profile. This fact is very important if we really want to make
a realistic prediction of the expected $\gamma$-ray flux. These flux
predictions have been already done for Draco using different models
for the DM density profiles
\cite{bergstrom,evans,colafrancesco,mambrini}. Nevertheless, in our
case, we compute these flux predictions for a cusp and a core DM
density profiles built from the latest stellar kinematic observations
together with a rigorous method of removal of interloper stars. This
computation represents by itself a recommendable update of the best DM
model for Draco, but as we will see it will be also useful to extract
some important conclusions concerning the possible uncertainties in
the absolute $\gamma$-ray flux coming from astrophysics.

Once we have obtained the flux profiles, we will use them to stress
the role of the Point Spread Function (PSF) of the
telescope. Including the PSF, which is directly related to the angular
resolution of the IACT, is essential to interpret correctly a possible
signal profile due to neutralino annihilation, not only for Draco but
also for any other target. In fact, we will show that, depending on
the PSF of the IACT, we could distinguish or not between different
models of the DM density profile using the observed flux profile. In
the case of the cusp and core DM density profiles that we use, it
could be impossible to discriminate between them if the PSF is not
good enough. It is worth mentioning that most of previous works in the
literature (except \cite{Prada04}) that calculated the expected flux
profiles in IACTs due to dark matter annihilation did not take into
account this important effect. Because of that, to emphasize the role
of the PSF constitutes also one of the goals of this work.

Finally, we present the DM detection prospects of Draco for some
current or planned experiments, i.e. MAGIC, GLAST and GAW. We carry
out the calculations under two different approaches: detection of the
gamma ray flux profile from the cusp and core DM models for Draco, and
detection of an excess signal in the direction of the dwarf galaxy. We
will show how the first approach could give us a lot of information
about the origin of the gamma ray flux profile, but it is slightly
harder to have success for this case compared to the second approach,
where even the PSF of the instrument is not essential and still we
could extract some important conclusions if we reached the required
sensitivity.

The paper is organized as follows. In Section \ref{sec2} we first
present all the equations necessary to properly calculate the expected
$\gamma$-ray flux in IACTs due to neutralino annihilation. Both the
particle physics and the astrophysics involved are described
carefully. For the particle physics we analyze neutralino annihilation
in the context of the minimal supersymmetric standard models (MSSM)
compatible with the present phenomenological and cosmological
constraints. In Section \ref{sec3} we show in detail the model that we
use for the DM distribution in Draco. In Section \ref{sec4} we
calculate the Draco flux predictions. We also stress the important
role of the PSF. In Section \ref{sec5}, the prospects to detect signal
due to neutralino annihilation in Draco are shown for some current or
planned experiments, i.e. MAGIC, GLAST and GAW. Conclusions are
finally given in Section \ref{sec6}.

\section{The $\gamma$-ray flux in IACT$s$} \label{sec2}

The expected total number of continuum $\gamma$-ray photons received
per unit time and per unit area in a telescope with an energy
threshold $E_{th}$ is given by the product of two factors:

\begin{equation}
F(E>E_{\rm th})=\frac{1}{4\pi} {f_{SUSY}} \cdot U(\Psi_0).
\label{eq1}
\end{equation}

\noindent where $\Psi_0$ represents the direction of observation
relative to the centre of the dark matter halo. The factor $f_{SUSY}$
includes all the particle physics, whereas the factor $U(\Psi_0)$
involves all the astrophysical properties (such as the dark matter
distribution and geometry considerations) and also accounts for the
beam smearing of the telescope.

\subsection{Particle physics: the $f_{SUSY}$ parameter}
In R-parity conserving supersymmetric theories, the lightest SUSY
particle (LSP) remains stable. The widely studied Minimal
Supersymmetric extension of the Standard Model (MSSM) can predict a
neutralino as the LSP with a relic density compatible with the WMAP
bounds. In this section we concentrate on the MSSM under the
assumption that the neutralino is the main component of the DM present
in the universe.  We consider its abundance inside the bounds $0.09 <
\Omega_\chi h^2 < 0.13$ as derived by fitting the $\Lambda$CDM model
to the WMAP data \cite{wmap}, although the possibility for other DM
candidates is not excluded (see Ref.~\cite{bertone, Munoz:2003gx} and
references therein).

The properties of the neutralinos in the MSSM are determined by its 
gaugino-higgsino composition:

\beq
\chi\equiv \chi^0_1=N_{11} \tilde{B}+N_{12} \tilde{W}^3
+N_{13} \tilde{H}^0_1+N_{13} \tilde{H}^0_2
\eeq

\noindent At leading order, neutralinos do not annihilate into
two-body final states containing photons. However, at one loop it is
possible to get processes such as \cite{Bergstrom:1997fh, Bern,
Ullio:1997ke}: \bea \chi+\chi &\rightarrow& \gamma \gamma \nonumber \\
\chi+\chi &\rightarrow& Z \gamma \nonumber \eea

\noindent with monochromatic outgoing photos of energies \beq
E_\gamma\sim m_\chi,\;\,\ E_\gamma\sim m_\chi- {m_Z^2\over{4 m_\chi}}
\eeq respectively. Also, the neutralino annihilation can produce a
continuum $\gamma$~--~ray spectrum from hadronization and subsequent
pion decay which can be dominate over the monochromatic $\gamma$'s.

\begin{figure}[!ht]
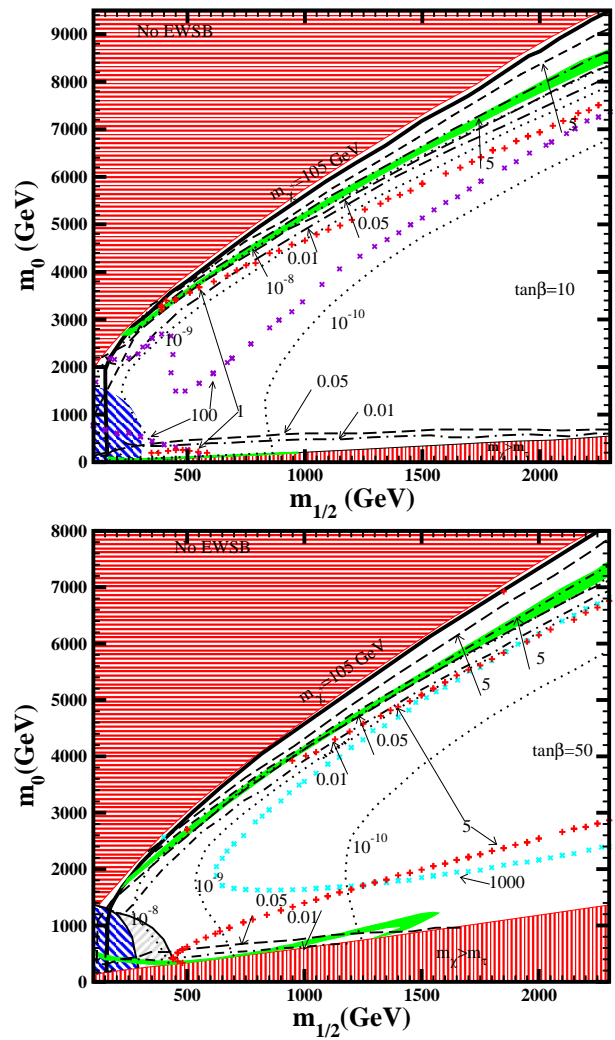

\begin{center}
\epsfig{file=m12m0_10.eps,width=8cm}
\epsfig{file=m12m0_50.eps,width=8cm}
\end{center}
\caption{Contours on the $m_0-m_{1/2}$ plane, the up and down
ruled areas are excluded by the not satisfaction of the EWSB (up) and
because $m_\chi>m_{\tilde{\tau}}$. The area below the upper thick
solid line satisfies the experimental bound on the chargino mass,
while the green shaded areas indicate the areas that predict
neutralino relic density on WMAP bounds. From left to right, the ruled
areas are excluded by the bounds on $m_h^0$ and $BR(b\rightarrow s
\gamma)$ respectively. The doted lines indicates the values of
$\sigma_{\chi p}$ in pb, the dash and dot-dash lines corresponds
respectively to $2 v \sigma_{\chi\chi\rightarrow \gamma \gamma}$ and $
v \sigma_{\chi\chi\rightarrow Z \gamma}$ in units of $10^{-29} cm^3
s^{-1}$. The line of crosses (plus) correspond to the continuum photon
production with $E_\gamma>1$~GeV ($E_\gamma>100$~GeV) also in units of $10^{-29} cm^3 s^{-1}$.}
\label{fig:areas}
\end{figure}

The lack of experimental evidence of supersymmetric particles 
leaves us with a number of undetermined parameters in the 
SUSY models. Therefore, the annihilation cross sections involved in both 
the computation of $\Omega_\chi h^2$ and $\gamma$ production can 
change orders of magnitude with neutralino mass and  
its gaugino--higgsino composition. To be specific we consider 
mSUGRA models, where the soft terms of the MSSM
are taken to be universal at the gauge unification scale
$M_{GUT}$. Under this assumption, the effective theory at energies
below $M_{GUT}$ depends on four parameters: the soft scalar mass
$m_0$, the soft gaugino mass $m_{1/2}$, the soft trilinear coupling
$A_0$, and the ratio of the Higgs vacuum expectation values,
$\tan\beta=\left<H_u^0\right>/\left<H_d^0\right>$. In addition, the
minimisation of the Higgs potential leaves undetermined the sign of
the Higgs mass parameter $\mu$.

To provide some specific values, we assume $A_0=0$, $\mu>0$ and two
values of $\tan\beta$, 10 and 50. With this two cases we can provide 
a qualitative picture of most relevant space of parameters and the 
constraints imposed by phenomenology. The impact of the size of 
$A_0$ or the sign of $\mu$ can be found in several studies of the parameter 
space of the MSSM (see for example 
Refs.~\cite{Cerdeno:2003yt,Gomez:2005nr,Gomez:2003cu,Stark} and references therein).

In Fig.~\ref{fig:areas} we displayed some lines of constant values of
$2\left< v \sigma_{\gamma \gamma}\right>$, $\left<v \sigma_{\gamma
Z}\right>$ and cross sections for continuum $\gamma-$ray emission on
the plane $m_0-m_{1/2}$ along with the constraints derived from the
lower bound on the mass of the lightest neutral Higgs,
$m_h^0=114.1$~GeV, chargino mass $m_{\tilde{\chi}^+}=103$~GeV and
BR($b\rightarrow s \gamma$), and the areas with $\Omega h^2$ on the
WMAP bounds. Also we provide the lines with constant values for the
elastic scattering $\chi$--proton, relevant for neutralino direct
detection. In the computation we used DarkSUSY \cite{Gondolo:2004sc}
combined with isasugra~\cite{Paige:2003mg} implementing the
phenomenological constraints as discussed in Ref.~\cite{Gomez:2005nr};
the estimation of BR($b\rightarrow s \gamma$) was performed using
Refs.  ~\cite{Belanger:2004yn,Gomez:2006uv}.

On the lower area consistent with WMAP the neutralino is Bino--like,
the relic density is satisfied mostly due to coannihilations
$\chi-\tilde{\tau}$ and in the case of $\tan\beta=50$ because of
annihilation $\chi-\chi$ through resonant channels. In this sector
$\left<2 v \sigma_{\gamma \gamma}\right>$ is dominant by a factor of
10 with respect to $\left<v \sigma_{\gamma Z}\right>$, however its
larger values lie in the areas constrained by the bounds on
$m_{\tilde{\chi}^+}$, $m_{h^0}$ and BR($b\rightarrow s \gamma$).

The higher area consistent with WMAP lies on the hyperbolic branch,
the neutralino is gaugino--higgsino mixed. The position of this region 
is very dependent on the mass of the top; we used $m_t=175$~GeV.

\begin{figure*}[!ht]
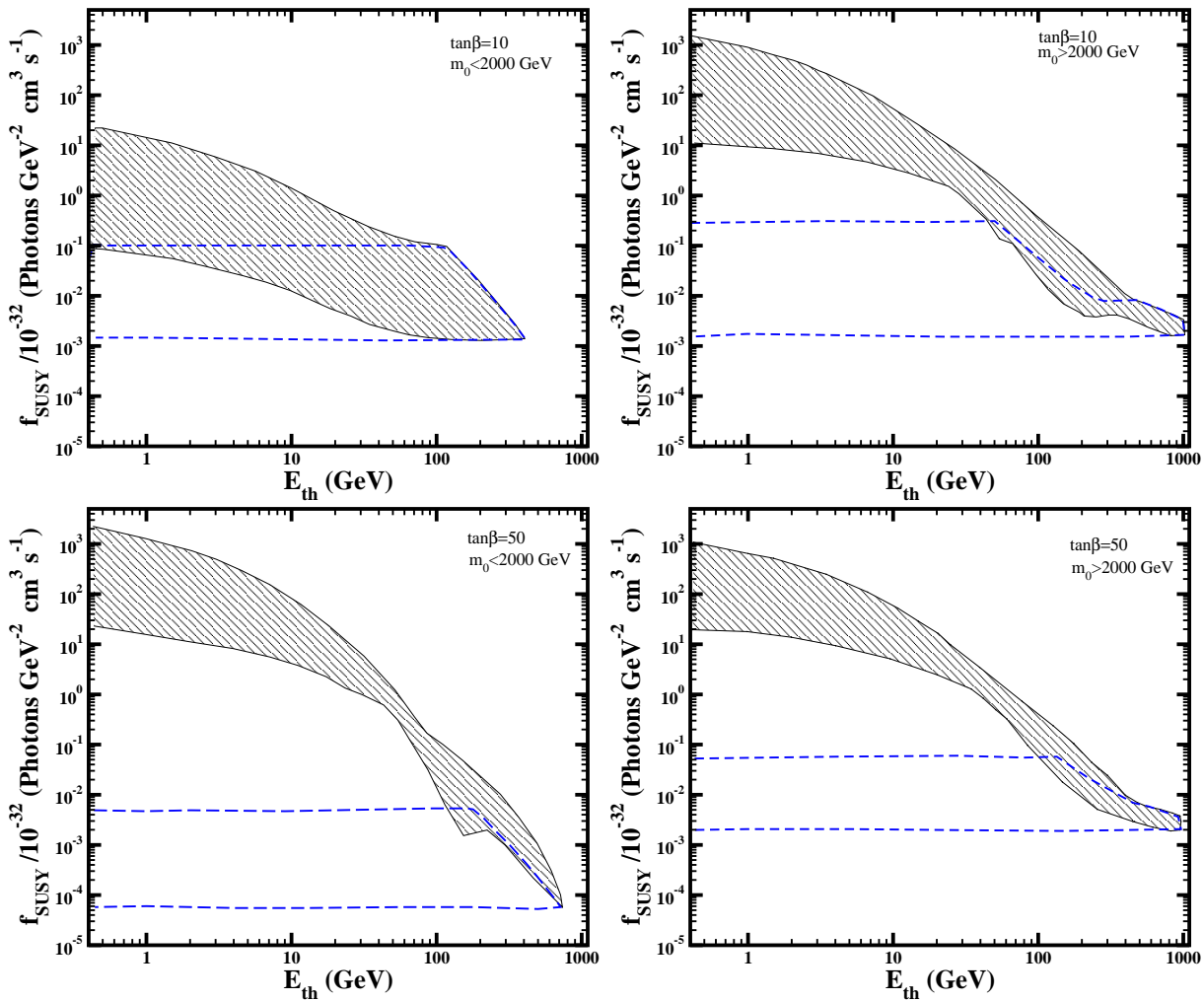

\begin{center}
\epsfig{file=fsusy_a10.eps,width=8cm}
\epsfig{file=fsusy_f10.eps,width=8cm}
\epsfig{file=fsusy_a50.eps,width=8cm}
\epsfig{file=fsusy_f50.eps,width=8cm}
\end{center}
\caption{Values of $f_{SUSY}$ respect $E_{th}$, for the points in
Fig.\ref{fig:areas} on the WMAP region and satisfying all the
phenomenological constraints. The ruled areas include the continuum
$\gamma-$ray emission with $E_\gamma>E_{th}$, while the ones limited
by thin lines correspond only to monochromatic channels.  The left
(right) panel corresponds to the upper (lower) allowed WMAP area in
Fig. 1.}
\label{fig:fsusy}
\end{figure*}

In Fig.~\ref{fig:fsusy} we present the values of $f_{SUSY}$ versus the
threshold energy of the detector, starting at $E_{th}=0.4$~GeV, for
neutralinos satisfying relic density and phenomenological
constraints. $f_{SUSY}$ is calculated as:

\begin{displaymath}
f_{SUSY}={\theta(E_{th}>m_{\chi})\cdot 2 \left<v \sigma_{\gamma
\gamma} \right>\over{2 m_{\chi}^2}}
\end{displaymath}
\begin{equation}
+{\theta(E_{th}>m_{\chi}-{m_Z^2\over{4m_\chi}})\cdot
\left<v \sigma_{\gamma Z}\right>+~k\left<v \sigma_{cont.}\right>
\over{2 m_{\chi}^2}}, 
\end{equation} 

\noindent where $\theta$ is the step function and $k$ the photon
multiplicity for each neutralino annihilation. We display in different
panels values of $f_{SUSY}$ corresponding to points on hyperbolic
branch ($m_0>2$~TeV as we can see in Fig.\ref{fig:areas}) and the
values corresponding to the $\chi-\tilde{\tau}$ coannihilation and
resonant annihilation areas ($m_0<2$~TeV). We can appreciate that on
the $\chi-\tilde{\tau}$ coannihilation area, $m_{\chi}$ has an upper
bound beyond which the relic density constraint is no longer satisfied
while on the hyperbolic branch no upper bound for $m_\chi$ is
reached. The largest values of $f_{SUSY}$ are not present since they
correspond to low values of $m_{1/2}$ suppressed by the bounds on
$m_{\tilde{\chi}^+}$, $m_{h^0}$ and $b\rightarrow s \gamma$. It is
interesting to remark that the higher values of $f_{SUSY}$ on the
constrained areas lie in which the $\sigma_{\chi-p}$ reaches values in
the range of direct detection experiments like Genius
\cite{Cerdeno:2003yt}.

\begin{figure}[!h]
\begin{center}
\epsfig{file=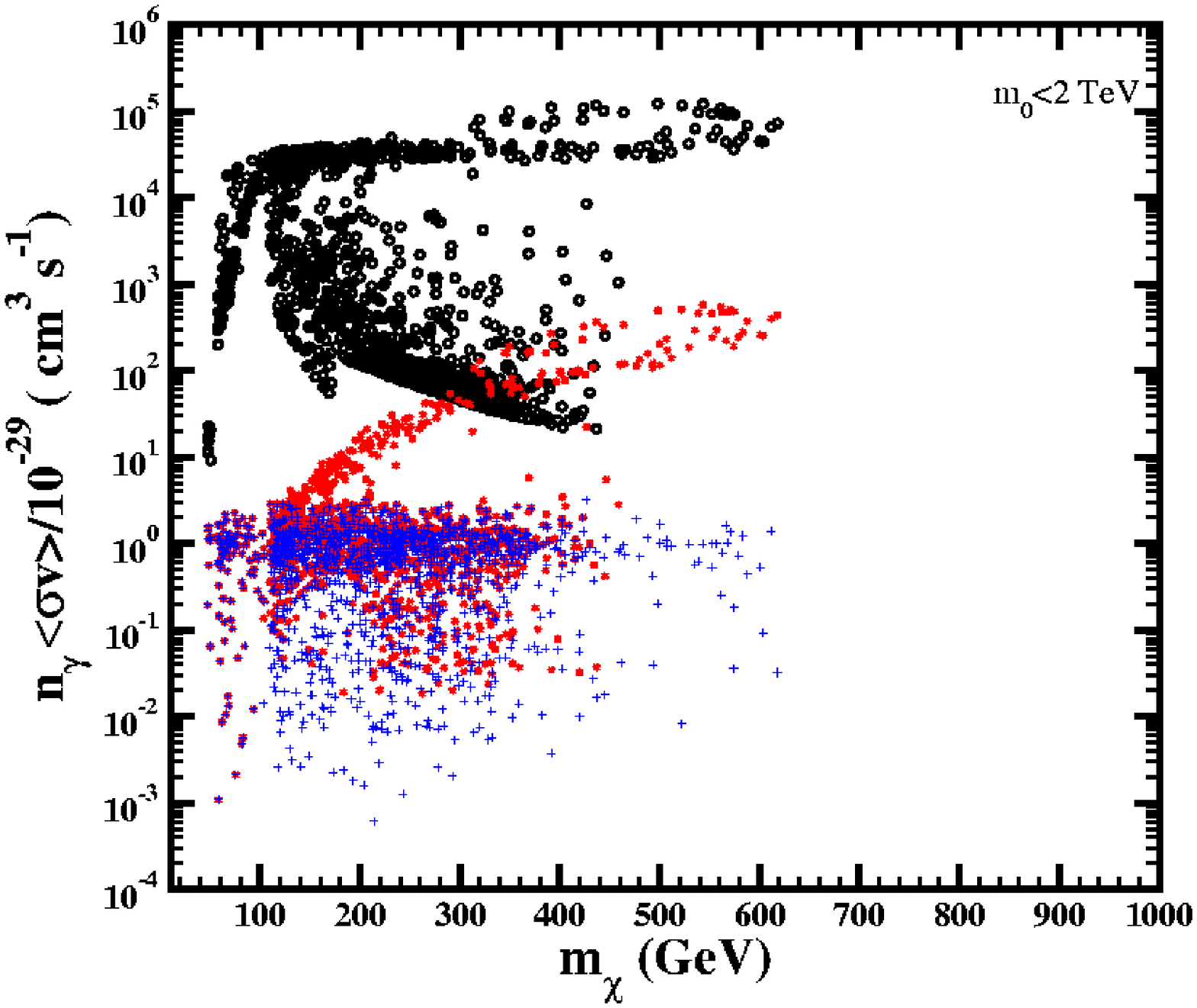,width=8cm}
\epsfig{file=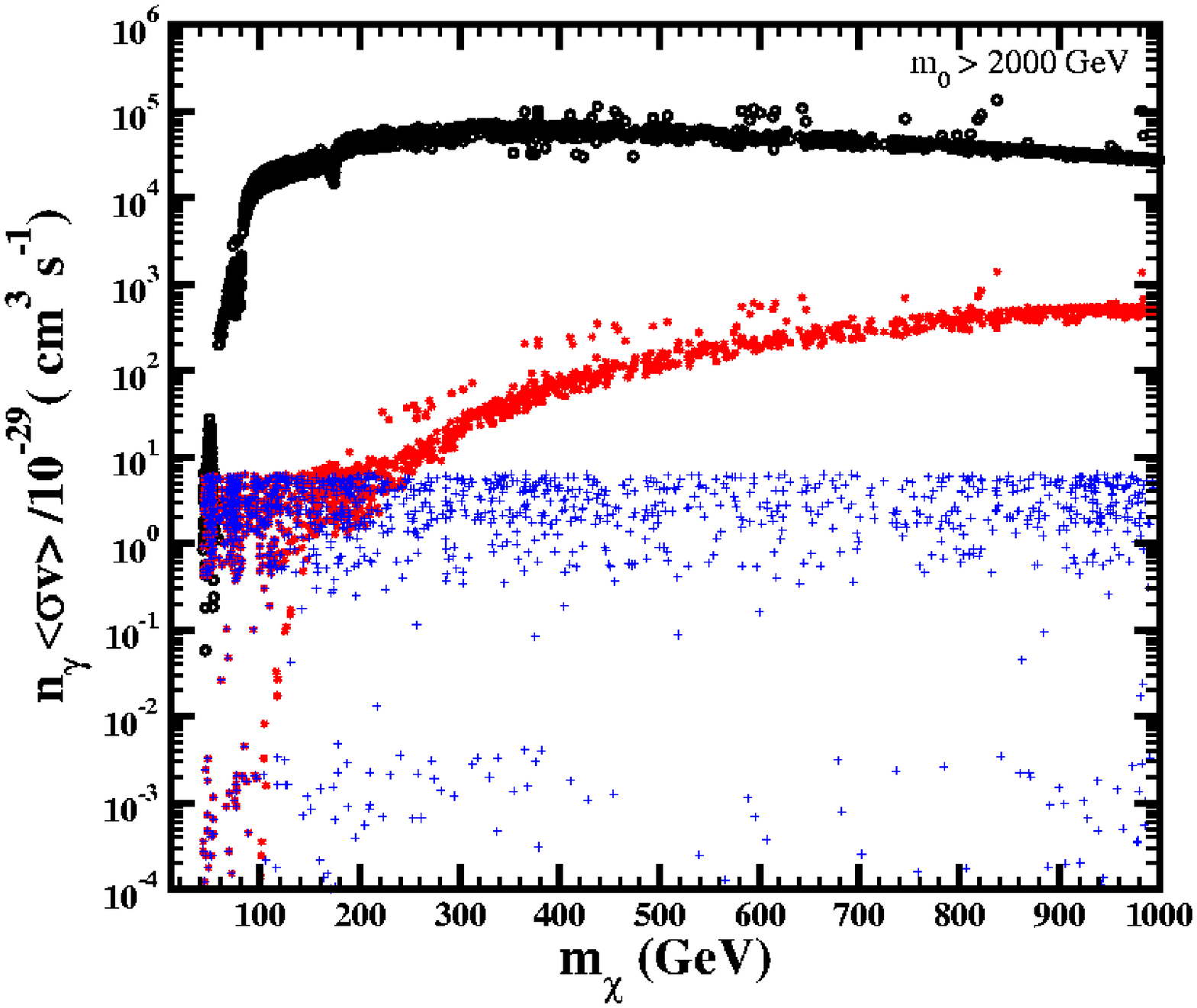,width=8cm}
\end{center}
\caption{Values of $n_\gamma <\sigma_{\chi \chi}v> $ in $cm^3/s$
including continuum emission for $E_\gamma>1$~GeV (circles),
$E_\gamma>100$~GeV (stars), and considering only the two monochromatic
channels (plus).}
\label{fig:scat}
\end{figure}

To show a more general sampling on  the parameter space, we present 
in Fig.~\ref{fig:scat} the $\gamma$-ray  production in 
neutralino annihilation versus $m_\chi$. Each panel contains about  1400 points 
satisfying the WMAP and phenomenological  constraints mentioned above. These 
points are selected from random scan over 600000 parameter on the ranges:

\bea
3<&\tan \beta&<60,\nonumber \\
50~\rm{GeV} < &m_{1/2}&<2300\, \rm{GeV} ,\nonumber \\
-3~m_0<&A_0&<3~m_0,\nonumber \\
50~\rm{GeV} < &m_0& <10~\rm{TeV}, 
\label{eq5}
\eea

\noindent and both signs of the $\mu$ term.

In the top panel of Fig.~\ref{fig:scat} we include models with $m_0
<2$~TeV. These points satisfy the relic density constraint due mostly
to $\tilde{\tau}-\chi$ coannihilations, and to a minor extend due to
resonances in the annihilation channels (for large values of
$\tan\beta$); $m_\chi$ remain below $700$~GeV in this panel. The
continuum emission dominates for $E_\gamma >1$~GeV, while for
$E_\gamma >100$~GeV we find points where the continuum production is
of the same order as the monochromatic $\gamma$'s. The bottom panel of
Fig.~\ref{fig:scat} contains models $m_0 > 2$~TeV. These correspond
mostly to the hyperbolic branch region; also we can appreciate that
larger values for $m_\chi$ are allowed in this area than in the top
panel. The points with larger cross sections in both panels
corresponds to the smaller values of the pseudo-scalar Higgs mass
$m_A$ and higher higgsino composition of the neutralino. These two
factors favour the fermion production in neutralino annihilation
channels. On the top panel, the larger cross sections correspond to
the larger values of $\tan\beta$ while on the right panel low values
for $m_A$ can be reached for any $\tan\beta$.

The mSUGRA results do not differ significantly that the ones 
obtained in the  more general MSSM models obtained by waiving 
the universality conditions on the soft terms as we will present 
in section \ref{sec5B}.

\subsection{Astrophysics: the $U(\Psi_0)$ parameter} \label{sec22}

All the astrophysical considerations are included in the expression
$U(\Psi_0)$ in Eq.(\ref{eq1}). This factor accounts for the dark
matter distribution, the geometry of the problem and also the beam
smearing of the IACT, i.e.

\begin{equation}
U(\Psi_0)=\int J(\Psi)B(\Omega)d\Omega
\label{eq1c}
\end{equation}

\noindent where $B(\Omega)d\Omega$ represents the beam smearing of the
telescope, commonly known as the Point Spread Function (PSF). The PSF
can be well approximated by a Gaussian:

\begin{equation}
B(\Omega) d\Omega  =  \exp\left[ -\frac{\theta^2}{2\sigma_t^2}\right] 
\sin\theta~ d\theta~d\phi
\label{eq3}
\end{equation}

\noindent with $\sigma_t$ the angular resolution of the IACT. It is
worth mentioning that there is some dependence of $\sigma_t$ with the
energy (see e.g. \cite{IACTparam}). However, for simplicity, we will
suppose this term to be constant. The PSF plays a very important role
in the way we will observe a possible DM signal in the
telescope. However, most of previous works in the literature did not
take into account its effect (except \cite{Prada04}; in \cite{profumo}
the PSF apparently was also used, although it is not mentioned in the
text (S. Profumo, private communication)). In Section \ref{sec4} we
will study in detail the importance of the PSF in the determination of
the gamma ray flux profile.

The $J(\Psi)$ factor of Eq.(\ref{eq1c}) represents the integral of the
line-of-sight of the square of the dark matter density along the
direction of observation $\Psi$:

\begin{equation}
J(\Psi) = \int_{l.o.s.} \rho_{dm}^2(r)~d\lambda =
\int_{\lambda_{min}}^{\lambda_{max}} \rho_{dm}^2[r(\lambda)]~d\lambda
\label{eq2}
\end{equation}

\noindent Here, $r$ represents the galactocentric distance, related to
the distance $\lambda$ to the Earth by:

\begin{equation}
r = \sqrt{\lambda^2+R_{\odot}^2-2~\lambda~R_{\odot} \cos \Psi}
\label{eq1a}
\end{equation}

\noindent where $R_{\odot}$ is the distance from the Earth to the
centre of the galactic halo, and $\Psi$ is related to the angles
$\theta$ and $\phi$ by the relation $\cos \Psi = \cos \psi_0 \cos
\theta + \sin \psi_0 \sin \theta \cos \phi$. The lower and
upper limits $\lambda_{min}$ and $\lambda_{max}$ in the
l.o.s. integration are given by $R_{\odot}\cos \psi \pm
\sqrt{r_t^2-R_{\odot}^2\sin^2 \psi}$, where $r_t$ is the tidal radius
of the dSph galaxy in this case.

\section{Dark matter distribution in Draco} \label{sec3}

In our modelling of Draco we used the sample of 207 Draco stars with
measured line-of-sight velocities originally considered as members by
\cite{wkeg}. In selecting these stars these authors relied on a simple
prescription going back to \cite{Yahil} and based on rejection of
stars with velocities exceeding $3\sigma_{\rm los}$ where $\sigma_{\rm
los}$ is the line-of-sight velocity dispersion of the
sample. \cite{lokas05} have shown that if all these 207 stars are used
to model Draco velocity distribution the resulting velocity moments
can be reproduced only by extremely extended mass distribution with
total mass of the order of a normal galaxy. Their arguments strongly
suggested that some of the stars may in fact be unbound and the simple
$3\sigma_{\rm los}$ rejection of stars is insufficient.

\begin{figure}
\centering \includegraphics[width=8cm]{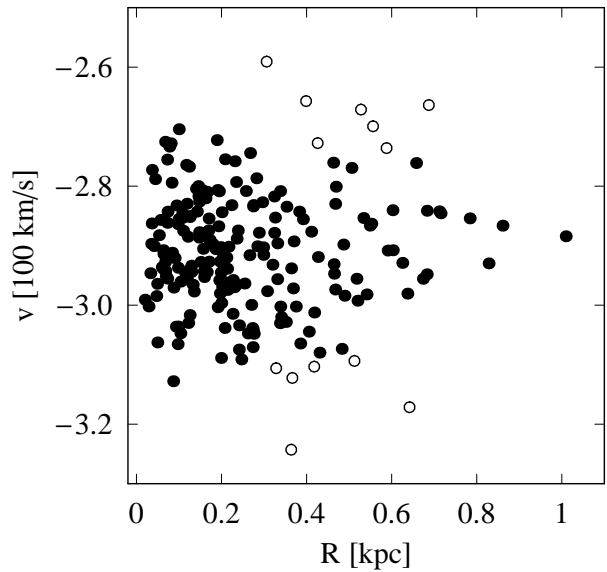}
  \caption{The line-of-sight velocities versus projected distances
from the galaxy centre for 207 stars from \cite{wkeg}. Open circles
mark the 13 stars rejected by our interloper removal procedure, filled
circles show the 194 ones accepted.} \label{stars194}
\end{figure}

Here we apply a rigorous method of removal of such interlopers
originally proposed by \cite{hartog} and applied to galaxy
clusters. The method relies on calculating the maximum velocity
available to the members of the object assuming that they are on
circular orbits or infalling into the structure. The method was shown
to be the most efficient among many methods of interloper removal
recently tested on cluster-size simulated dark matter haloes by
\cite{wojtak}. Its applicability and efficiency in the case of dSph
galaxies was demonstrated by \cite{klim}. Fig.~\ref{stars194} shows
the results of the application of this procedure to Draco. The 207
stars shown in the plot are divided into those iteratively rejected by
the procedure (open circles) and those accepted at the final iteration
(filled circles).

\begin{figure}
\centering
  \includegraphics[width=8cm]{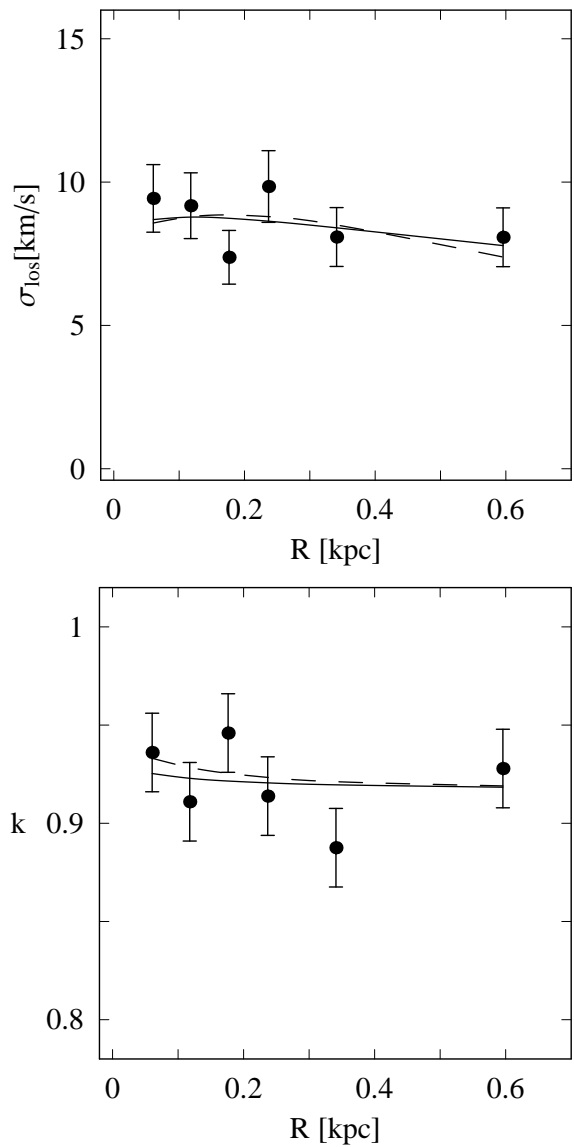}
  \caption{The line-of-sight velocity dispersion (upper panel) and
kurtosis variable $k$ (lower panel) calculated for the sample of 194
stars with 32-33 stars per bin. The lines show the best-fitting
solutions of the Jeans equations for the DM profile with a cusp (solid
lines) and a core (dashed lines).} \label{dk194}
\end{figure}

The final sample with 194 stars is different from any of the three
considered by \cite{lokas05} therefore we repeat their analysis here
for this new selection. Our analysis is exactly the same, except that
in the calculation of the velocity moments we use 32-33 stars per bin
instead of about 40 and we consider a DM profile with a core in
addition to the cusp one. The profiles of the line-of-sight velocity
moments, dispersion and kurtosis, obtained for the new sample are
shown in Fig.~\ref{dk194}. The kurtosis was expressed in terms of the
variable $k = [\log(3 K/2.7)]^{1/10}$ where $K$ is the standard
kurtosis estimator. We assumed that the DM distribution in Draco can
be approximated by
\begin{equation}    \label{kazantzidis}
    \rho_{\rm d}(r) = C r^{-\alpha} {\rm exp} \left( -\frac{r}{r_{\rm
    b}} \right)
\end{equation}
proposed by \cite{kmmds}, which was found to fit the density
distribution of a simulated dwarf dark matter halo stripped during its
evolution in the potential of a giant galaxy. In the same work it was
found that the halo, which initially had a NFW distribution, preserves
the cusp in the inner part (so that $\alpha=1$ fits the final remnant
very well) but develops an exponential cut-off in the outer
parts. Here we will consider two cases, the profile with a cusp
$\alpha=1$ and a core $\alpha=0$. It remains to be investigated which
scenarios could lead to such core profiles.

\begin{figure}
\centering
  \includegraphics[width=7.2cm]{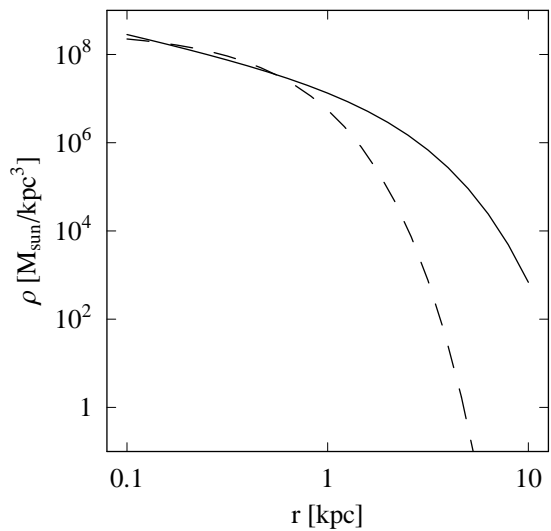}
  \caption{The best-fitting DM density profiles for Draco with a cusp
(solid line) and a core (dashed line).}  \label{density194}
\end{figure} 

The best-fitting solutions to the Jeans equations (see \cite{lokas05})
for two component models with dark matter profiles given by
(\ref{kazantzidis}) are plotted in Fig.~\ref{dk194} as solid lines in
the case of the cusp profile and dashed lines for the core. The
best-fitting parameters of the two models are listed in
Table~\ref{parameters}, where $M_{\rm D}/M_{\rm S}$ is the ratio of
the total dark matter mass to total stellar mass, $r_{\rm b}/R_{\rm
S}$ is the break radius of equation (\ref{kazantzidis}) in units of
the S\'ersic radius of the stars and $\beta$ is the anisotropy
parameter of the stellar orbits.

\begin{table}
\caption{\label{parameters}Best-fitting parameters of the two-component models for the
DM profiles with a cusp ($\alpha=1$) and a core ($\alpha=0$) obtained
from joint fitting of velocity dispersion and kurtosis profiles shown
in Fig.~\ref{dk194}.  The last column gives the goodness of fit
measure $\chi^2/N$.}
\begin{ruledtabular}
    \begin{tabular}{ccccccdcc}
      profile &  & $M_{\rm D}/M_{\rm S}$ & & $r_{\rm b}/R_{\rm S}$ & & \beta  & & $\chi^2/N$  \\
      \hline
      cusp    &  & 830  & &  7.0 & & -0.1 & &  8.8/9   \\
      core    &  & 185  & &  1.4 & & 0.06  &  &  9.5/9   \\
    \end{tabular}
\end{ruledtabular}
\end{table}

Fig.~\ref{density194} shows the best-fitting dark matter density
profiles in the case of the cusp (solid line) and the core (dashed
line). As we can see, both density profiles are similar up to about 1
kpc, where they are constrained by the data. The reason for very
different values of the break radius $r_{\rm b}$ in both cases is the
following. The kurtosis is sensitive mainly to anisotropy and it
forces $\beta$ to be close to zero in both cases. However, to
reproduce the velocity dispersion profile with $\beta \approx 0$ the
density profile has to be steep enough. In the case of the core it
means that the exponential cut-off has to occur for rather low radii,
which is what we see in the fit. The cusp profile does not need to
steepen the profile so much so it is much more extended and its total
mass is much larger.

\section{Draco gamma ray flux profiles} \label{sec4}

In order to compute the expected gamma flux, we need to calculate the
value of the ``astrophysical factor'', $U(\psi_0)$, given in
Eq.(\ref{eq1}) and presented in detail in Section \ref{sec22}. We
calculated it for the core ($\alpha=0$) and cusp ($\alpha=1$) density
profiles given by Eq.(\ref{kazantzidis}) using the parameters listed
in Table \ref{parameters2} (that were deduced from those given in
Table \ref{parameters} and where we used $R_S=7.3$ arcmin for Draco,
following \cite{odenkirchen}). $R_{\odot}$ was set to 80 kpc, as
derived from an analysis on the basis of wide-field CCD photometry of
resolved stars in Draco \cite{aparicio}. For the tidal radius we used
a value of 7 kpc as given by \cite{evans} and derived from the Roche
criterion supposing an isothermal profile for the Milky
Way. Nevertheless, this value depends strongly on the profile used for
the Milky Way and Draco, e.g. a value of 1.6 kpc is found when a NFW
DM density profile is used for both galaxies \cite{evans}. It is worth
mentioning, however, that we computed $J(\Psi)$ for different values
of $r_t$ and we checked that the difference between choosing $r_t =
1.6$ kpc and $r_t = 7$ kpc is less than 5\% for $\Psi=0.5\grado$, and
still less than 10\% for $\Psi=1\grado$. Therefore, for the case of
Draco, any value $r_t \gtrsim 1.5$ leads to robust and very similar
results.

\begin{table}
  \caption{\label{parameters2}Values of $C$ and $r_b$ for a cusp and
  a core DM density profile given by Eq.(\ref{kazantzidis}), as
  deduced from those parameters listed in Table \ref{parameters}.}
  \begin{ruledtabular}
    \begin{tabular}{ccccc}
      profile &  & $C$ & & $r_b~(kpc)$ \\
      \hline
      cusp    &  & 3.1 x $10^7$~M$_{\odot}$/kpc$^2$ & &  1.189  \\
      core    &  & 3.6 x $10^8$~M$_{\odot}$/kpc$^3$ & &  0.238  \\
    \end{tabular}
  \end{ruledtabular}
\end{table}

There is another issue that we will have to take into account in order
to calculate $U(\psi_0)$. If we integrate the square DM density along
the line of sight using a cusp DM density profile, we will obtain
divergences at angles $\psi_0 \rightarrow 0$ (clearly there will not
be any problem for core profiles). This can be solved by introducing a
small constant DM core in the very centre of the DM halo. In
particular, the radius $r_{cut}$ at which the self annihilation rate
$t_l \sim (<\sigma_{ann}v>n_{\chi}~r_{cut})^{-1}$ equals the dynamical
time of the halo $t_{dyn} \sim (G~\overline{\rho})^{-1/2}$, where
$\overline{\rho}$ is the mean halo density and $n_{\chi}$ is the
neutralino number density, is usually taken as the radius of this
constant density core \cite{fornengo}. For the NFW DM density profile
this value for $r_{cut}$ is of the order of $10^{-13}-10^{-14}$
kpc. For steeper DM density profiles (such as the compressed NFW or
the Moore profile) a value of $r_{cut} \sim 10^{-8}$ kpc is obtained.
We used a value of $10^{-8}$ kpc in all our computations. We must
note that $r_{cut}$ represents a lower limit concerning the acceptable
values for this parameter, so the obtained fluxes should be taken as
upper bounds.

Once we have calculated $U(\psi_0)$, we will need also to take a value
for the $f_{SUSY}$ parameter in order to obtain the absolute flux due
to neutralino annihilation (see Eq.~\ref{eq1}). We chose a value of
$f_{SUSY} = 10^{-33}~ph~GeV^{-2}~cm^{3}~s^{-1}$ in all our
computations for a typical $E_{th} \sim 100$ GeV of the IACT. In the
framework of MSSM, this value corresponds to one of the most
optimistic values that is possible to adopt for $f_{SUSY}$ according
to Fig.~\ref{fig:fsusy} for the two different values of tan $\beta$
presented.

\begin{figure}
\centering
  \includegraphics[width=8.5cm]{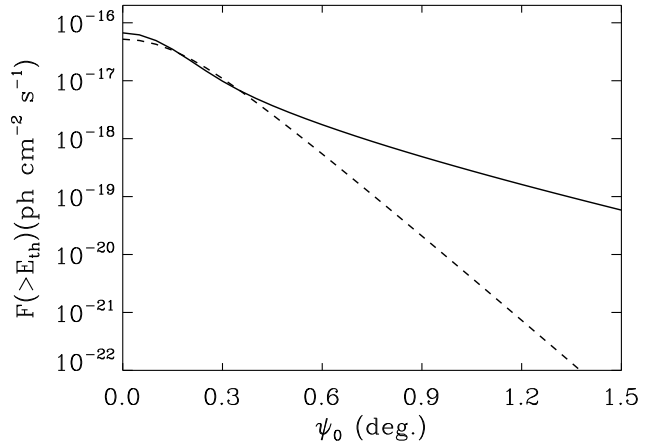}
  \caption{Draco flux predictions for the core (dashed line) and cusp
  (solid line) density profiles, computed for a typical IACT with
  E$_{th}=100$ GeV and a PSF$=0.1\grado$. We used $f_{SUSY} =
  10^{-33}~ph~GeV^{-2}~cm^{3}~s^{-1}$ (see text for details).}
  \label{cuspcorepsf}
\end{figure}

The resulting $\gamma$-ray flux profiles for Draco are plotted in
Figure \ref{cuspcorepsf}, where we used a PSF with
$\sigma_t=0.1\grado$ (to simplify the notation, hereafter we will use
PSF$=0.1\grado$ to refer to a PSF with $\sigma_t=0.1\grado$). This
value of 0.1$\grado$ is the typical value for an IACT like MAGIC or
HESS. It is important to note that the core and a cusp density
profiles would be distinguishable thanks to a different and
characteristic shape of the flux profile in each case.

\begin{figure}[!h]
\begin{center}
\includegraphics[width=8.5cm]{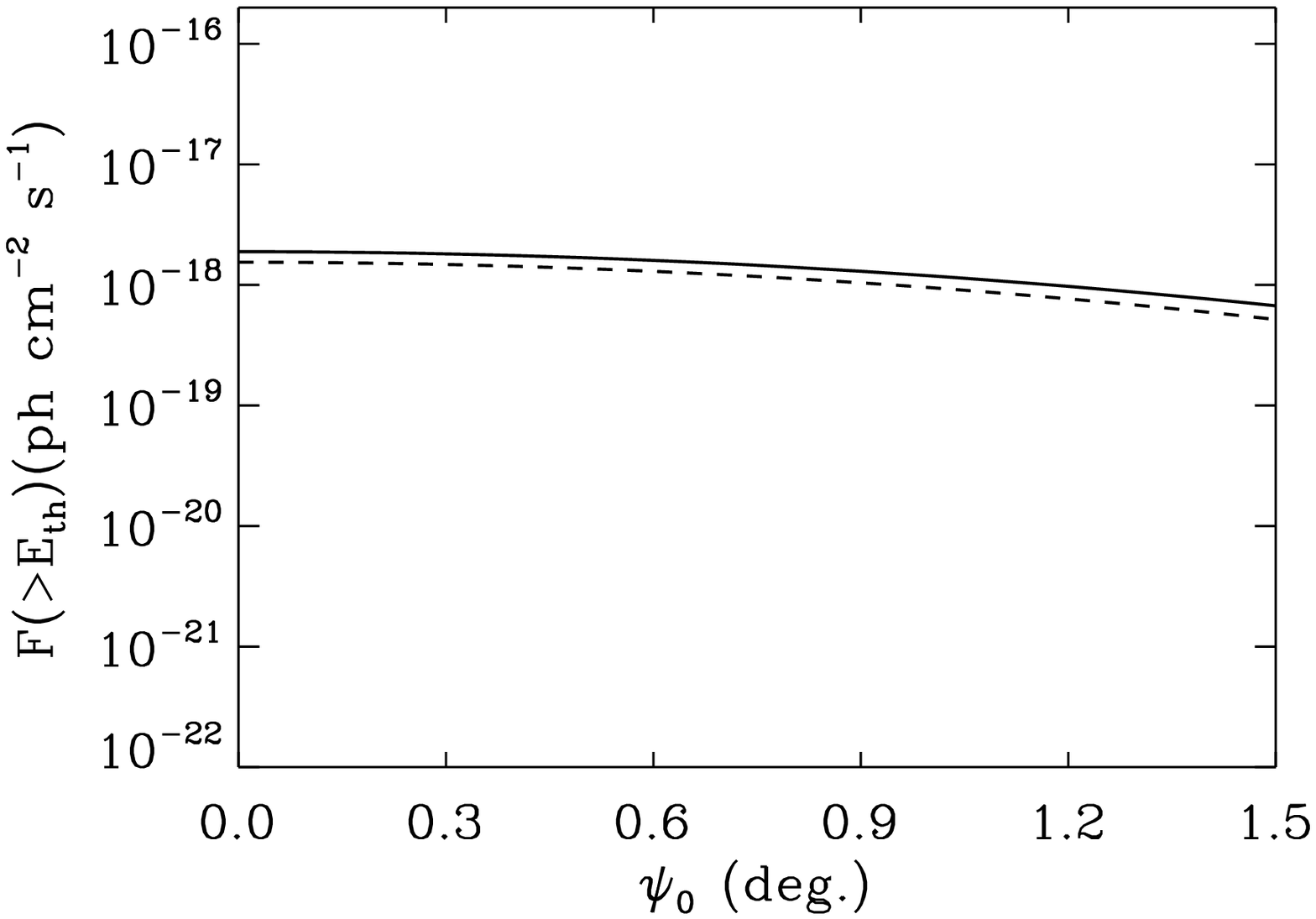}
\includegraphics[width=8.5cm]{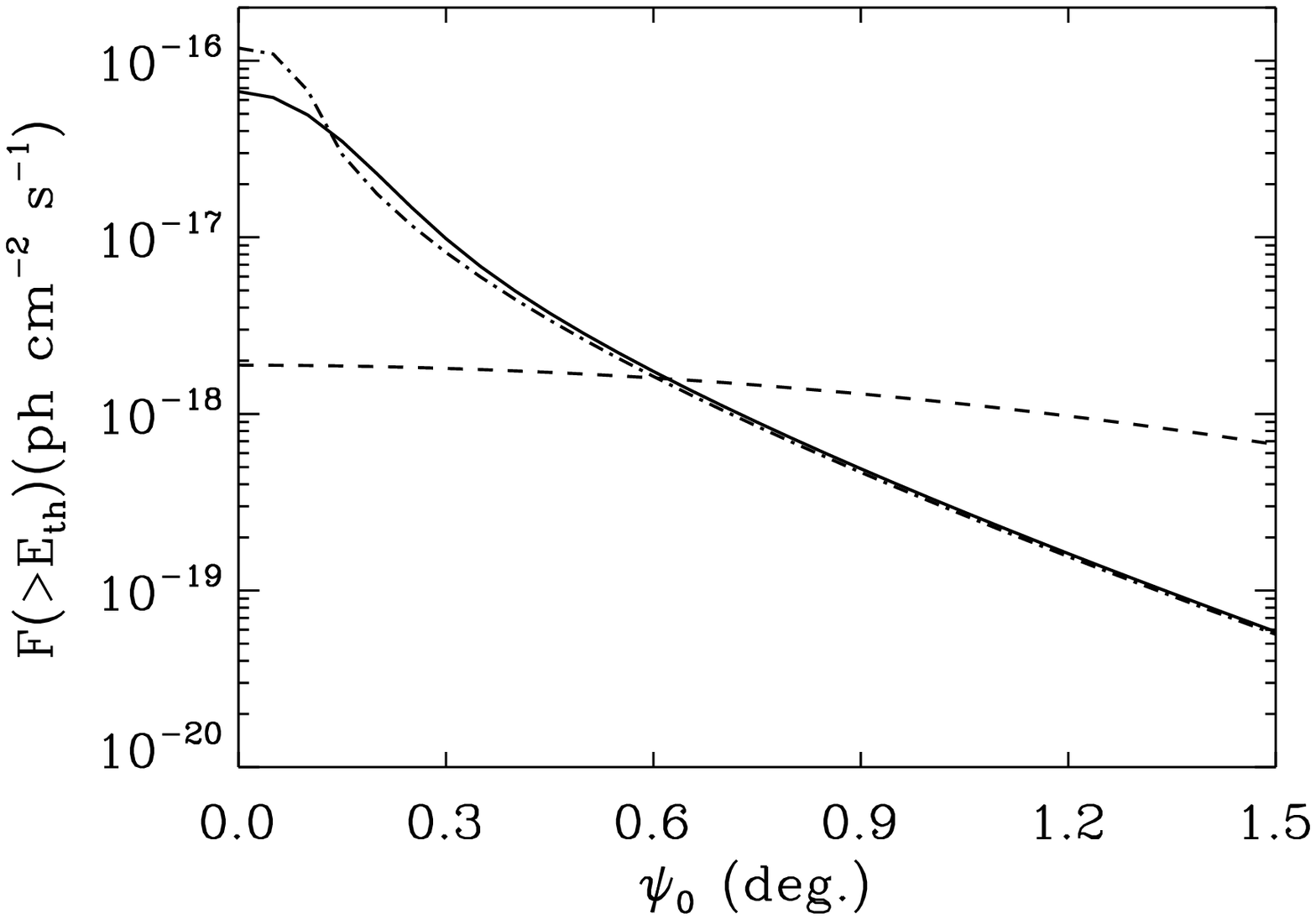}
\end{center}
\caption{Top panel: Draco flux predictions for the core (dashed line)
  and cusp (solid line) density profiles, computed using a
  PSF$=1\grado$. Bottom panel: Draco flux predictions for the cusp
  density profile using two different PSFs. Solid line corresponds to
  PSF=$0.1\grado$ and dashed line to PSF=$1\grado$. The flux profile
  computing without PSF is also shown for comparison (dot-dashed
  line).}
\label{psf}
\end{figure}

To illustrate the PSF effect on the shape of the observed flux profile
with IACTs, in the top panel of Figure \ref{psf} we show the same as
in Fig.~\ref{cuspcorepsf}, but here for a PSF$=1\grado$. As we can
see, although we have different DM density profiles, a worse telescope
resolution makes both resulting flux profiles for a core and a cusp
indistinguishable. We may think that we could distinguish them from
the value of the absolute flux. However, the difference in the
absolute flux between both DM density profiles is very small and in
practice the distinction would be impossible. There are many
uncertainties in this absolute flux coming from the particle
physics. $f_{SUSY}$ may be very different from the most optimistic
case assumed here, since it could vary more than three orders of
magnitude for this SUSY model (see Fig.~\ref{fig:fsusy}). On the other
hand, the uncertainty in flux due to the DM density profile to be core
or cusp is negligible at least in the inner 0.5 degrees.

Concerning the effect of the PSF given the same DM density profile, a
worse telescope resolution flattens the flux profile as expected. It can be
clearly seen in the bottom panel of Figure \ref{psf}, where we plot the Draco
$\gamma$-ray flux predictions only for the cusp density profile but using two
different values of the PSF ($0.1\grado$ and $1\grado$), and where we
plot also the same flux profile computed without PSF for
comparison. 

A good example to show the real importance of the telescope resolution
can be found around the controversy generated in the wake of the Draco
$\gamma$-ray excess reported by the CACTUS collaboration in 2005
\cite{marleau}. At this moment, it seems clear that this excess was
not real and was probably due to instrumental and trigger-related
issues. However, concerning our line of work and always just with the
intention of clarifying the role of the PSF, we must mention the
results shown in \cite{profumo}. There, the CACTUS data were
superimposed on different flux profiles (each of them related to
possible models of DM density profiles for Draco) in Figure 2. These
flux profiles were computed using an angular resolution of
$0.1\grado$, whereas the CACTUS PSF is quite worse than that (around
0.3$\grado$ for the Crab and probably worse for Draco
\cite{profumo,cactus}).  Looking at that figure (despite the
authors' indication to take it with care) one may come to the
conclusion that a core profile seems to be the most adequate DM
density profile for Draco, as opposed to the cusp profile. However, it
would be more appropriate to make this comparison between the CACTUS
data and the flux profiles using in both cases the same PSF of the
experiment. Doing so and taking into account the PSF effect properly,
it would be difficult to use the CACTUS data to discriminate between
different models for the DM density profile as described in
\cite{profumo}, since all of the resultant flux profiles would have
essentially the same shape. Only the absolute flux could give us a
clue to make this distinction possible, but as already mentioned there
are too many uncertainties related to an absolute value to be able to
extract solid conclusions.

\section{Detection prospects for some current or planned experiments} \label{sec5}
\subsection{Flux profile detection}
Would it be possible to detect a signal due to neutralino annihilation
in Draco using present or planned IACTs and satellite-based gamma ray
experiments? Although there are many uncertainties concerning both
particle physics and astrophysical issues, as already pointed out, it
is possible (and necessary) to make some calculations. These
calculations will allow us to estimate at least the order of magnitude
of the flux that we could expect in our telescopes, and will help us
learn which instrument is, in principle, the best positioned and
optimised to detect a possible signal from Draco.

Draco is located in the northern hemisphere, more precisely at
declination $58\grado$. Because of that, and regarding currently
operating IACTs, MAGIC \cite{magic} and VERITAS \cite{veritas} are the
best options thanks to their geographical position. Since both
experiments have comparable sensitivity and PSF, we will focus only on
MAGIC. For this IACT, with an energy threshold around 50 GeV for
zenith observations, Draco could be observed $61\grado$ above the
horizon, so an energy threshold $\sim 100$ GeV seems to be still
possible at that altitude.

Concerning GLAST \cite{glast}, this satellite-based experiment is
designed for making observations of celestial gamma-ray sources in the
energy band extending from $\sim 10$ MeV to $300$ GeV, which is
complementary to the one for MAGIC. Moreover, it will have a PSF $\sim
0.1\grado$ at 10 GeV \cite{webglast}, which will make this instrument
very competitive also in DM searches. GLAST is planned to be launched
at the end of 2007.

\begin{figure}[!h]
\begin{center}
\includegraphics[width=8.5cm]{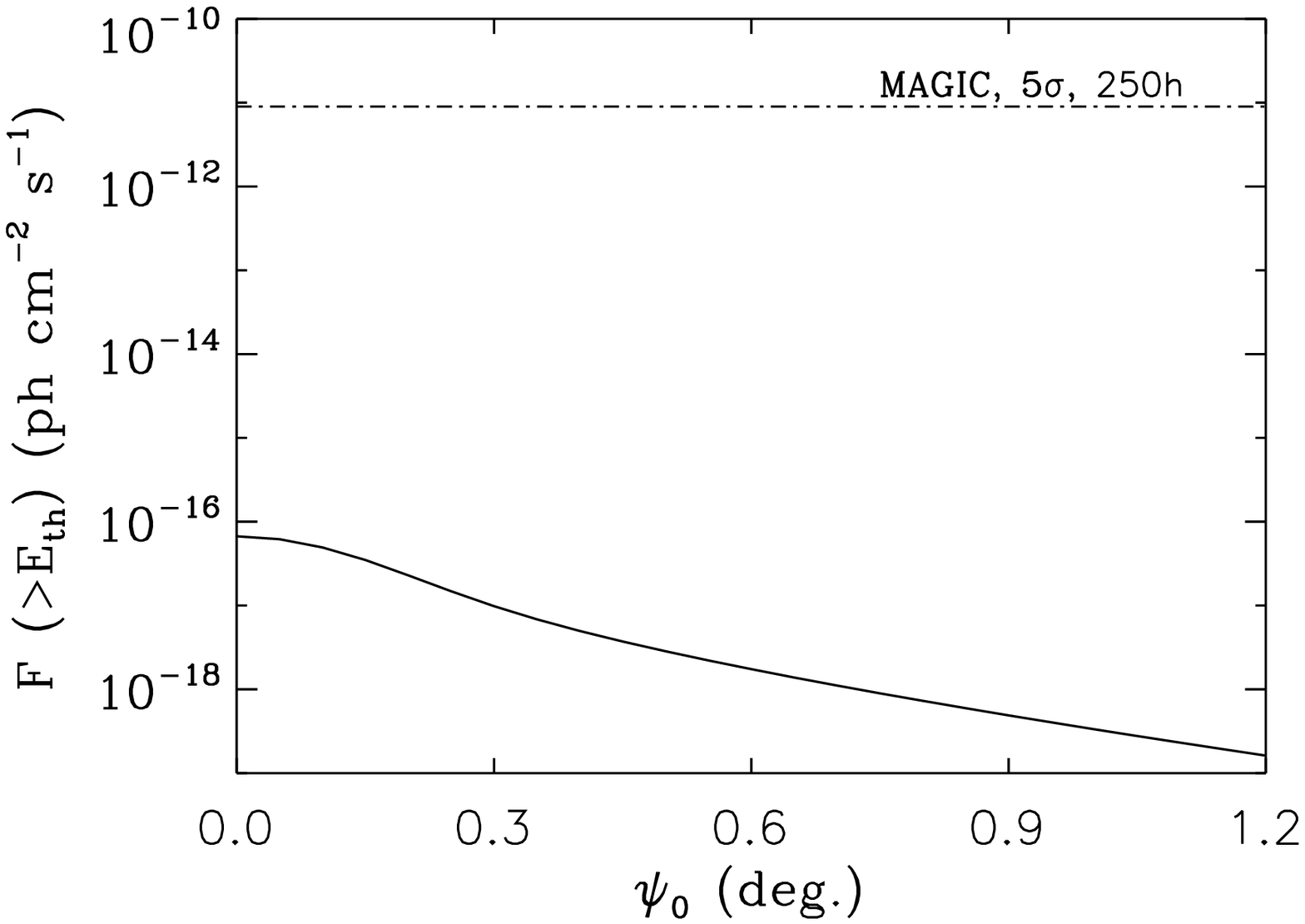}
\includegraphics[width=8.5cm]{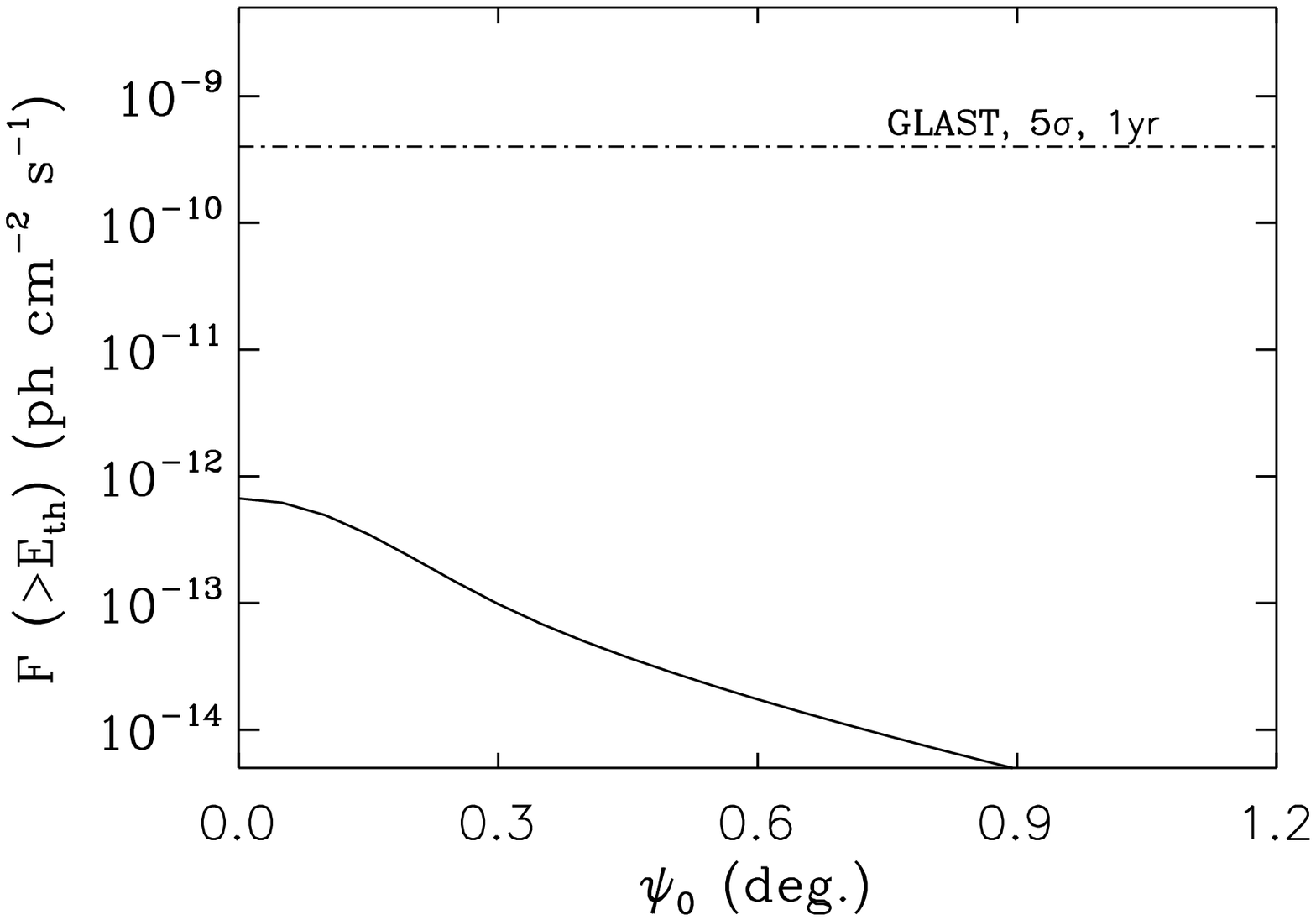}
\end{center}
\caption{Draco flux profile detection prospects for MAGIC (top panel)
  and GLAST (bottom panel). The flux profile (solid line) corresponds
  to the cusp density profile given in Table \ref{parameters2} and
  using a PSF=$0.1\grado$. The sensitivity lines for both instruments
  were computed for a 5$\sigma$ detection level and 250 hours (MAGIC)
  / 1 year (GLAST) of integration time. These figures should be taken
  as the most optimistic case, since we used the most optimistic
  scenario from particle physics in each case. We adopt E$_{th}=100$
  GeV for MAGIC and E$_{th}=1$ GeV for GLAST. See text for details.}
\label{prospects}
\end{figure}

The flux profiles detection prospects for both gamma-ray experiments
can be seen in Figure \ref{prospects}, where the lines of sensitivity
for MAGIC and GLAST are superimposed on the flux profile computed
using the cusp DM density profile for Draco as given by
Eq.(\ref{kazantzidis}) together with the parameters listed in Table
\ref{parameters2}. For the case of MAGIC, the sensitivity line
represents 250 hours of observation time and a $5\sigma$ detection
level. As pointed out before, this curve should be comparable and
valid for VERITAS experiment as well. Concerning the predicted
sensitivity for GLAST, it was calculated by the GLAST team
\cite{webglast} for 1 year of observation and a $5\sigma$ detection
level. Values of $f_{SUSY}=10^{-33}$~ph~GeV$^{-2}$~cm$^{3}$~s$^{-1}$ at 100
GeV (MAGIC) and $f_{SUSY}=1 \cdot 10^{-29}$~ph~GeV$^{-2}$~cm$^{3}$~s$^{-1}$
at 1 GeV (GLAST) were chosen to convert the values calculated from the
astrophysical factor $U(\Psi_0)$ to flux. These values of $f_{SUSY}$
correspond to the most optimistic values compatible with that shown in
Fig.~\ref{fig:fsusy}. We chose E$_{th}=1$ GeV for GLAST to be sure
that we will have a good angular resolution for the arriving photons
and also to avoid some uncertainties at low energies in the response
of the GLAST LAT detector as pointed out in \cite{bertone06}.

From Figure \ref{prospects} we can see that the detection of the gamma
ray flux profile due to neutralino annihilation in Draco with the
MAGIC telescope seems to be impossible (at least for the DM density
profiles and the particle physics model used here), since we would
need roughly five orders of magnitude more sensitivity than that
reached by this instrument in a reasonable time to have at least one
opportunity to detect a possible signal coming from DM
annihilation. For the case of GLAST we obtain more or less the same,
although in this case the expectations are substantially better and we
would need three orders of magnitude more sensitivity. We must note
that this is the most optimistic scenario, so even leaving space to
take into account the large uncertainties coming from the particle
physics it seems very hard to have any chance of detection. In fact,
in case of adopting a pessimistic value for $f_{SUSY}$, the
$\gamma$-ray flux profile shown in Fig.~\ref{prospects} could decrease
in more than three orders of magnitude easily.

There are some issues concerning a possible detection of DM
annihilation not only in Draco but also in any other possible target
that should be taken into account at this moment. In the case of a
positive detection, this signal would be diffuse (i.e. no point-like
source) for an instrument with a PSF good enough. This fact and the
characteristic spectrum of the source represent the best clues to
distinguish between a $\gamma$-ray signal due to neutralino
annihilation from other astrophysical sources. In the case of Draco,
however, we are very far from obtaining any of these clues, so Draco
does not seem a good target for DM searches for current or planned
$\gamma$-ray experiments (MAGIC-II, for example, will have a factor of
2 more sensitivity than the single MAGIC telescope \cite{magic2},
still clearly insufficient, and the same improvement in sensitivity is
reached in the case of the GLAST satellite if we observe 5 years
instead of 1).

\subsection{Excess signal detection} \label{sec5B}

Although it is strongly recommendable to try to detect the gamma ray
flux profile due to neutralino annihilation, we could also search only
for a gamma ray excess signal in the direction of Draco with a DM
origin. In this case the prospects should be somewhat better, since we
are only interested in flux detectability (i.e. no interested in
observing an extended emission or not; no discrimination between
different flux profiles). Draco is around $2\grado$ in the sky (the
whole DM halo), and one MAGIC pointing is 4$\grado$x 4$\grado$, so the
whole galaxy is within one of these MAGIC pointings. This means that,
to calculate the prospects of an excess signal detection for MAGIC, we
must sum the gamma ray fluxes coming from all the neutralino
annihilations that occur in the entire halo of the dSph. The same is
valid for the case of GLAST, since one of the main objectives of this
mission will be to survey the whole sky, so Draco as a whole will be
observed. Nevertheless, most of the $\gamma$-ray flux due to DM
annihilation comes from the inner regions of the dwarf, so it would be
better to integrate the flux and make all the calculations only for
those regions. Otherwise, if we integrate up to very large angles from
the center, we would be increasing the noise in a large amount without
practically increasing the $\gamma$-ray signal, since the gamma ray
flux profile decreases very rapidly from the center to the outskirts.

We would like here to take the opportunity to mention GAW \cite{gaw},
which is a R\&D path-finder experiment, still under development, for
wide field $\gamma$-ray astronomy. GAW will operate above 0.7 TeV and
will have a PSF $\sim 0.2\grado$. It will consist of three identical
telescopes working in stereoscopic mode (80m side). The main goal of
GAW is to test the feasibility of a new generation of IACTs, which
join high sensitivity with large field of view (24$\grado$ x
24$\grado$). GAW is planned to be located at Calar Alto Observatory
(Spain) and a first part of the array should be completed and start to
operate during 2008. It is a collaborative effort of research
Institutes in Italy, Portugal, and Spain. We will also present some
calculations concerning the possibility to observe a $\gamma$-ray
signal in the direction of Draco by GAW, just to illustrate the
capabilities of the instrument. Nevertheless, the main advantages of
GAW will point to other directions, e.g. the possibility to survey a
large portion of the sky in a reasonable time above 0.7 TeV.

In Table \ref{excesssignal} we show the prospects of an excess signal
detection ($5\sigma$ level) for MAGIC, GLAST and GAW and for the case
of the cusp DM density profile. Both the integrated absolute fluxes
for Draco and the values given for the sensitivities (i.e. the minimum
detectable flux, $F_{min}$) refer to the inner 0.5$\grado$ of the
galaxy (although the total size of Draco in the sky is $\sim$
2$\grado$), just to improve the signal to noise ratio as explained
above. The integrated absolute flux for Draco is not the same for the
three experiments, since the $f_{SUSY}$ parameter depends on the
energy threshold of each instrument, which is different (we chose 100
GeV for MAGIC, 1 GeV for GLAST and 700 GeV for GAW). We did the
calculations for the most optimistic case (i.e. the highest value of
$f_{SUSY}$ that we could adopt in the MSSM scenario following
Fig.~\ref{fig:fsusy}, and given the energy threshold of each
telescope) and the most pessimistic one (the lowest $f_{SUSY}$).

\begin{table}
  \caption{\label{excesssignal}Prospects of an excess signal detection
  for MAGIC, GLAST and GAW. Concerning the integrated flux for Draco,
  $F_{Draco}$, the most optimistic and pessimistic values are given in
  the form $F_{Draco,min}$ - $F_{Draco,max}$. $F_{min}$ represents the
  minimum detectable flux for each instrument. Both the integrated
  absolute fluxes for Draco and the values given for the sensitivities
  (i.e. the minimum detectable flux, $F_{min}$) refer to the inner
  0.5$\grado$ of the galaxy (although its size is $\sim$ 2$\grado$),
  just to improve the signal to noise ratio as explained in the text.}
  \begin{ruledtabular}
    \begin{tabular}{ccccc}
       &  & $F_{Draco}$ (ph~cm$^{-2}$~s$^{-1}$) & & $F_{min}$ (ph~cm$^{-2}$~s$^{-1}$) \\
      \hline
      MAGIC  &  & 4.0 x 10$^{-18}$ - 4.0 x 10$^{-16}$ & &  2.0 x 10$^{-11}$ (250 h)\\
      GLAST  &  & 4.0 x 10$^{-16}$ - 4.0 x 10$^{-12}$ & &  4.0 x 10$^{-10}$ (1 year) \\
      GAW    &  & 2.8 x 10$^{-19}$ - 2.8 x 10$^{-17}$ & &  2.2 x 10$^{-12}$ (250 h) \\
    \end{tabular}
  \end{ruledtabular}
\end{table}

According to Table \ref{excesssignal}, an excess signal detection
would be impossible with any of these three instruments, since even in
the most optimistic cases they do not reach the required sensitivity
by far. If we compare these numbers with the maximum fluxes shown in
Fig.\ref{prospects}, we can see that the enhancement, although
relevant (roughly a factor of 4 in the best case), is still very
insufficient for a successful detection.

Moreover, from the results shown in Table \ref{excesssignal} it
becomes really hard to extract any useful results that would help us
understand at least a bit better the problem of the dark
matter. Indeed, if we observe Draco with MAGIC or GLAST and we do not
detect any excess signal, we will not be able to put any useful
constraints on the particle physics involved. None of the actual
allowed values for $f_{SUSY}$ could be rejected if we find no
detection. This makes Draco even less attractive for those current
gamma ray experiments that try to find DM traces.

\begin{figure}[!h]
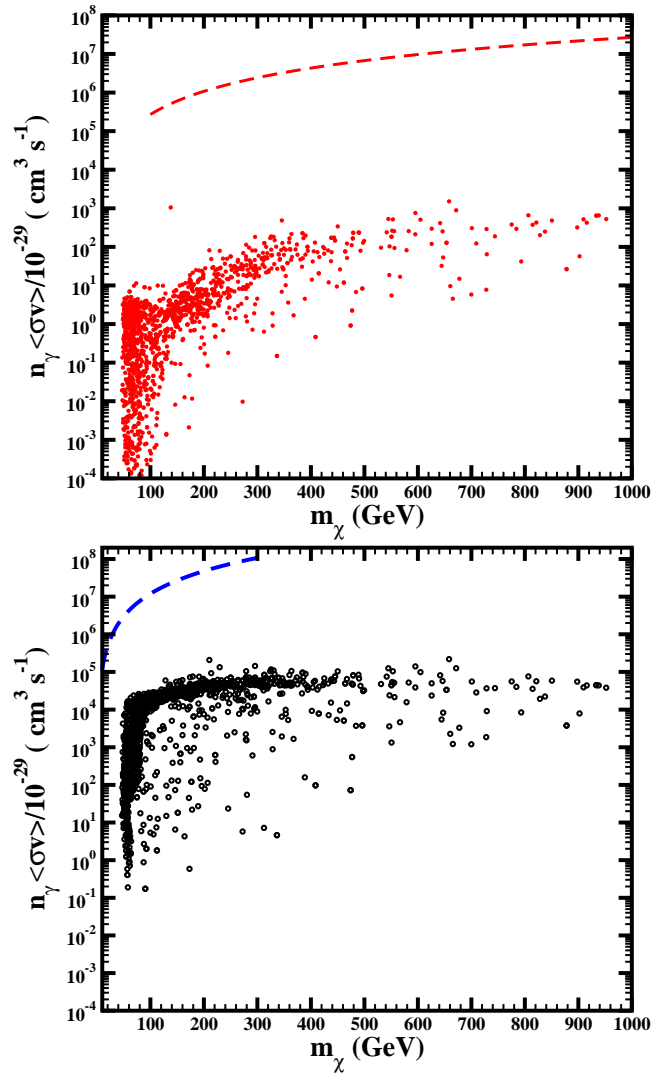

\begin{center}
\includegraphics[width=8.5cm]{rnd_magic.eps}
\includegraphics[width=8.5cm]{rnd_glast.eps}
\end{center}
\caption{Exclusion limits for MAGIC (upper panel) and GLAST (bottom
panel), for continuum $\gamma$-ray emission above 100 GeV (MAGIC) and
1 GeV (GLAST). The lines represent for each instrument the minimum
detectable $n_{\gamma}<\sigma v>$ adopting the cusp DM density profile
given in Table \ref{parameters2} for Draco. We used 250 hours (1 year)
of integration time for MAGIC (GLAST) and a 5$\sigma$ detection
level. Below the lines, the SUSY models (points) do not yield a flux
high enough for a successful detection. According to these plots,
neither MAGIC nor GLAST may put any useful constraints at least to
this particular particle physics model.}
\label{scatter}
\end{figure}

This can be clearly seen in Fig.~\ref{scatter}, where we show the
parts of the SUSY parameter space that will be detectable by MAGIC and
GLAST for the case of Draco and adopting the cusp DM density profile
presented in Section \ref{sec3}. The points represent MSSM models, in
contrast with Fig.~\ref{fig:scat} the mSUGRA condition of
Eq.~\ref{eq5} has been waived such that models with random sfermion
masses below 10 TeV and gaugino masses below 3 TeV are included (we
considered equal soft terms for the first two generations to avoid
contradiction with flavor violating observables). In the case of MAGIC
we only plot those values of $n_{\gamma}<\sigma v>$ computed for
$E_{\gamma}>100$ GeV (which is the MAGIC $E_{th}$), and for GLAST only
those ones computed for $E_{\gamma}>1$ GeV (GLAST $E_{th}$). We must
note that for each case we include in the same figure all the points
no matter the value of $m_0$ (a distinction was done in
Fig.~\ref{fig:scat}). Also, the points related to the monochromatic
channels are not shown in both figures, since their values are
negligible comparing to those coming from the continuum emission and
therefore they are not relevant here.

The limit between the detectable and the non-detectable areas is given
by the dashed lines, so that those SUSY points that lie above these
lines yield a detectable flux and the points below them are not
accessible to observation. As already pointed above, it is clear that
the constraints imposed by both experiments to this particular
particle physics model (MSSM) are very relaxed and in fact the
detection lines do not reach in any case any of the SUSY models by
more than two orders of magnitude in the best case. Again, as in
Fig.~\ref{prospects}, the prospects for GLAST are somewhat better than
for MAGIC, but they are still clearly insufficient to extract relevant
results or conclusions.

\section{Conclusions} \label{sec6}

In this work we focused on the possibility to detect a signal coming
from neutralino annihilation in the Draco dwarf. This galaxy, a
satellite of the Milky Way, represents one of the best suitable
candidates to search for dark matter outside our galaxy, since it is
near and it has probably more observational constraints than any other
known dark matter dominated system. This fact becomes crucial when we
want to make realistic predictions of the expected observed
$\gamma$-ray flux due to neutralino annihilation.

Draco is a dwarf galaxy tidally stripped by the Milky Way, so it seems
preferable to build a model for the mass distribution that takes into
account this important fact. Using this more appropriate model for
Draco, we have obtained the $\gamma$-ray flux profiles for the case of
a cusp and a core DM density profiles (both scenarios are equally
valid according to the observations). To do that, we first estimated
the best-fitting parameters for each density profile by adjusting the
solutions of the Jeans equations to velocity moments obtained for the
Draco stellar sample cleaned by a rigorous method of interloper
removal. Apart from this recommendable and useful update on the best
DM model for Draco, one important conclusion can be extracted
concerning the absolute $\gamma$-flux: for both cusp and core DM
density profiles, the flux values that we obtain are very similar for
the inner region of the dwarf, i.e. where we have the largest flux
values and signal detection would be easier.

There is, however, a way to distinguish between a core and a cusp DM
density profile. The crucial points concerning this issue are the
sensitivity and the PSF of the telescope. If the telescope resolution
is good enough (and we reach the required sensitivity) a distinction
between both cusp and core models may be possible thanks to the shape
of the flux profile in each case. However, if the PSF of the
instrument is poor, its effect could make it impossible to
discriminate between different flux profiles, i.e. different models of
the DM density profile. In any case, to be sure that the signal is due
to neutralino annihilation, we will need to have a PSF good enough to
be able to resolve the source (i.e. we will need to detect with a good
resolution at least a portion of the flux profile large enough so we
can conclude that it belongs to neutralino annihilation). This fact
together with a characteristic spectrum are the unequivocal traces of
a DM source. Both issues are of course totally valid not only for
Draco but also for any other target, and therefore they should be
always taken into account.

We may think that for Draco the effect of the PSF is not especially important, since current experiments do not seem to reach even the required sensitivity to detect a $\gamma$-ray signal coming from the center of the dSph. Indeed, the limiting factor here is the expected flux and not the PSF of the telescope. Nevertheless, we should account for the PSF of the instrument in any case, since we do not know \textrm{a priori} the expected gamma flux and we will not have a realistic prediction unless we include it in our calculations. Only then we will know the exact shape of the flux profile as could be observed by the telescope and therefore we can evaluate the real chances of detection. Furthermore, we must note that the inclusion of the PSF corresponds to a more general analysis not only valid for Draco but also for any target (the case the Milky Way for example where the role of the PSF is critical to understand the origin of the gamma emission in its center).

Some estimations concerning flux profile detection prospects for the
MAGIC and GLAST experiments have been also shown. According to these
calculations, even in the most optimistic scenario (i.e. with the
highest values of $f_{SUSY}$ allowed for the MSSM particle physics
model adopted here) a $\gamma$-ray signal detection from Draco seems
to be really hard at least for the current IACTs and for the GLAST
satellite. In addition, it would be really difficult to improve our
expectations for Draco, since for example larger integration times
will not improve drastically the sensitivity lines of these
instruments up to the level allowing successful detection (we need
around five and three orders of magnitude more sensitivity for success
with MAGIC and GLAST respectively). The uncertainties coming from
astrophysics, as already mentioned, are negligible compared to those
coming from the particle physics, so either by choosing a cusp or a
core DM density profile for Draco (or even modifying this for a
steeper one) we will not be able to increase the $\gamma$-ray flux up
to the level where we can expect a signal detection.

We also explored the prospects of an excess signal detection (i.e. we
are not interested in the shape of the gamma ray flux profile, only in
detectability) for MAGIC and GLAST, but we reached the same strongly
negative conclusions as those obtained for the flux profile
detection. Both instruments are very far from the required sensitivity
so it seems that they do not have any chance of successful
detection. But even more, in the case of an (expected) unsuccessful
detection we would not be able to put any useful constraints on the
particle physics involved (the uncertainties from different dark
matter density profiles and other astrophysical considerations are
negligible compared to this). None of the actual allowed values for
$f_{SUSY}$ could be rejected if we find no detection. This makes Draco
even less attractive for those current gamma ray experiments that try
to find DM traces.

From these negative results it seems that, at least for current
experiments, Draco does not represent a good target for DM searches
(although some important enhancements mechanisms in the $\gamma$-ray
signal coming from neutralino annihilation in Draco have been proposed
in \cite{colafrancesco}). If we want to find unequivocal traces of
non-barionic dark matter in the Universe, an effort should be done in
order to find other more promising DM targets. In this context, it is interesting to note that the number of known dSph galaxies satellites of the Milky Way has increased considerably in the last few years, doing possible that some of them may offer more optimistic detection prospects than Draco itself \cite{strigari}. A quantitative study should be done in this direction. The search of DM subhalos in the solar neighborhood (see e.g. \cite{diemand}) or exploring the IMBHs scenario \cite{imbh} may represent also other challenging possibilities, although these one more speculative for the moment.

Finally, it is worth mentioning that IACTs that join a large field of
view with a high sensitivity will represent the future in this field
and will provide a next step in DM searches. In this context, the
Cherenkov Telescope Array (CTA) project will be specially important,
with a threshold energy well below 100 GeV, a wide spectral coverage
up to 100 TeV and a factor of 5-10 more sensitivity than present
experiments in the mentioned range. Also GAW, a R\&D experiment under
development, with an energy threshold $\sim$ 700 GeV and a 24$\grado$
x 24$\grado$ field of view, will constitute another complementary
attempt in the same direction, although under a different approach.

\begin{acknowledgments}
 M.A.S.C. acknowledges the support of an I3P-CSIC fellowship in
Granada. M.A.S.C. and F.P. also acknowledge the support of the Spanish
AYA2005-07789 grant. E.L.{\L}. and R.W. are grateful for the
hospitality of the Instituto de Astrofis\'ica de Andaluc\'ia during
their visit. This work was partially supported by the Polish Ministry
of Science and Higher Education under grant 1P03D02726 and the
Polish-Spanish exchange program of CSIC/PAN. M.E.G. acknowledges
support from the 'Consejer\'{\i}a de Educaci\'on de la Junta de
Andaluc\'{\i}a', the Spanish DGICYT under contracts BFM2003-01266,
FPA2006-13825 and European Network for Theoretical Astroparticle
Physics (ENTApP), member of ILIAS, EC contract number
RII-CT-2004-506222.
\end{acknowledgments}

\end{document}